# THE SPECTRAL INVESTIGATION OF SEVEN HII REGIONS IN KAZARIAN GALAXIES

V. Zh. ADIBEKYAN

According to SDSS DR5 spectra the spectrophotometric investigations of seven HII regions of six Kazarian galaxies are conducted. The abundances of heavy elements and helium and also quantity of ionizing stars and star formation rate are determined. The oxygen abundance 12+log(O/H) lies in the range 7.94 ÷ 8.35. The mean log(S/O), log(Ar/O) and log(Ne/O) abundance ratios are equal to: -1.63, -2.37 and -0.78, respectively. The log(N/O) abundance ratio of investigated HII regions is in the interval -0.63 ÷ -1.37. They occupy the same area in the diagram N/O–O/H as the high-excitation HII regions. Most likely, the ages of investigated HII regions are larger than 100-300 Myr, required for the enrichment in nitrogen by intermediate-mass stars. The star formation rate is one order as in HII regions in spiral and irregular galaxies, and is in the interval 0.05÷0.81 $M_\odot \text{year}^{-1}$.

## 1. Introduction

The study of chemical composition of galaxies is fundamental for our understanding of the chemical evolution of galaxies. Emission-line galaxies provide an easy way to determine the element abundances, from an analysis of the radiation from their HII regions in the visible domain. HII regions are gaseous systems whose luminosity originates in the short-wavelength radiation from groups of early O-B stars (OB associations [1]). Both model calculations and empirical methods are used to determine the chemical composition of the HII regions. These determinations are considered more reliable if the electron temperatures can be measured directly, using the [OIII]λ4363/[OIII]λ5007 line ratio. Then, the ionic abundances can be derived directly from the strengths of the emission lines. (direct methods; [2,3]). The total abundance of a given element is given by the sum of abundances of all its ions. In practice, however, not all of the ions that are present can be observed in visible spectra (the only favorable case is that of oxygen) and one must correct for unseen ions using ionization correction factors (ICF). Unfortunately, in HII regions the oxygen-rich emission lines which are sensitive to temperature, e.g., O[III]λ4363, are often too faint for the electron temperatures to be determined directly. Then empirical methods (often referred to as "strong line methods") are used [4-6]. Empirical relations for determining the abundance of the chemical elements are obtained from the intensities of the strong emission lines.

The properties of HII regions in galaxies depend on the integral properties and Hubble type of the galaxies [7,8]. It has also been shown [7] that the abundances of chemical elements in the HII regions in spiral galaxies

Erevan State University, Armenia; e-mail: adbvardan@rambler.ru

depend strongly on their location within a galaxy (radial distance from the galactic center), unlike in irregular galaxies [9].

Oxygen and, probably, carbon are produced primarily by massive stars [10]. The origin of nitrogen is more controversial [11,12]. In nearby galaxies with star formation and an oxygen abundance 12+log(O/H)<7.6, the N/O ratio is almost constant, with very low dispersion around the mean value of log(N/O)=−1.6, which has led to the idea [11] that in these galaxies the nitrogen has been produced by massive stars as a primordial element. Elsewhere [12], it has been suggested that in galaxies with the same abundance 12+log(O/H)<7.6, the nitrogen was produced by intermediate stars with masses of 4-8 $M_\odot$ as a primordial element. The dependence of N/O on O/H is of crucial for understanding the origin of the nitrogen and the history of star formation in these objects. It has been shown [13] that in HII regions of spiral galaxies with high metallicity (12+log(O/H)>8.4), the ratio N/O increases linearly with the oxygen abundance, while in HII regions with low metallicity (12+log(O/H)< 8.4), this ratio is independent of the oxygen abundance. It has also been shown [13] that the HII regions located farthest from the centers of the parent galaxies (spiral galaxies) have the same values of N/O as dwarf galaxies with low metallicities.

In the present work the stellar population and the chemical composition of seven HII regions in Kazarian galaxies are investigated. Primary data of these galaxies are resulted in [14,15].

The spectra and the methods for analyzing them are discussed in Section 2. The abundances of chemical elements are derived in Section 3. The methods for determining the stellar population are presented in Section 4 and the conclusions are given in Section 5.

## 2. The spectra and and the analyzing methods

The SDSS DR5 (Sloan Digital Sky Survey, Data Release 5) is an enormous database of astronomical data containing direct images of about 217 million objects and spectra of about 1050000 objects (of which 675000 are galaxies) [16]. For studies of the Kazarian galaxies using the spectra from SDSS DR5, they have been identified with objects in the SDSS. Of the 94 identified Kazarian galaxies for which direct images exist, 65 have spectra in the spectral range 3800-9200 Å (including spectra of individual HII regions in spiral and irregular galaxies). From these we have chosen 7 HII regions in six Kazarian galaxies, in whose spectra the [OIII]λ4363 line was observed. Spectra of the HII regions (flux-calibrated) were taken from the web page of the SDSS at the http://www.sdss.org/dr5.

TABLE 1. Parameters of the HII Regions and their Host Galaxies

|  | Kaz 429 | Kaz 453 | Kaz 459 | Kaz 460(I) | Kaz 460(II) | Kaz528 | Kaz 530 |
|---|---|---|---|---|---|---|---|
| Morph. type. | Sd | Irr | Sd | Sc | Sc | Sdm | Irr |
| $M_g$ (gal.) | -17.8 | -19.8 | -18.4 | -19.4 | -19.4 | -18 | -18.5 |
| $M_g$ (HII) | -14.8 | -19.8 | -16.5 | -14.9 | -14.1 | -17.6 | -18.5 |
| $R/R_{25}$ | 0.44 |  | 0.6 | 0.59 | 0.42 | 0.75 |  |
| z | 0.0304 | 0.0471 | 0.0128 | 0.0105 | 0.0105 | 0.0178 | 0.022 |

Table 1 lists the parameters of the HII regions and their parent galaxies. The magnitudes of the HII regions and their parent galaxies in g(4684 A) band, which were used to calculate the absolute magnitudes, are taken from SDSS DR5. The values of R25 (the major semiaxis of the parent galaxy at the 25 mag arcsec$^{-2}$ isophotal level) are taken from the Hyperleda database, which is available at the http://leda.univ-lyon.fr/, and were used to determine the relative radial distances of the HII regions from the centers of the parent galaxies.

Direct images and spectra of the HII regions are shown in Figs. 1 and 2, respectively.

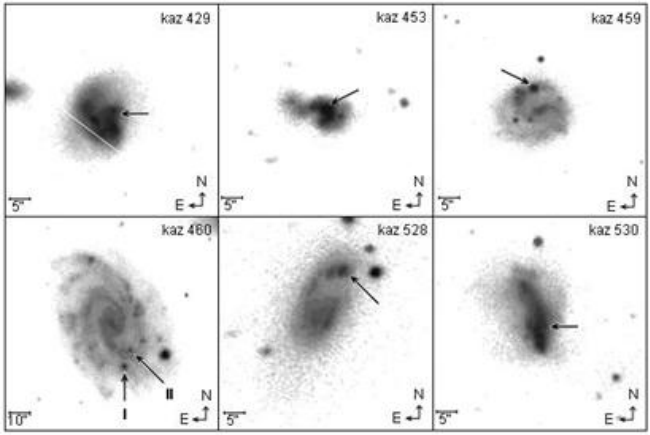

Fig. 1. Direct images of the HII regions and their parent galaxies from SDSS DR5. The HII regions are indicated by arrows and the two HII regions in Kaz 460 are numbered.

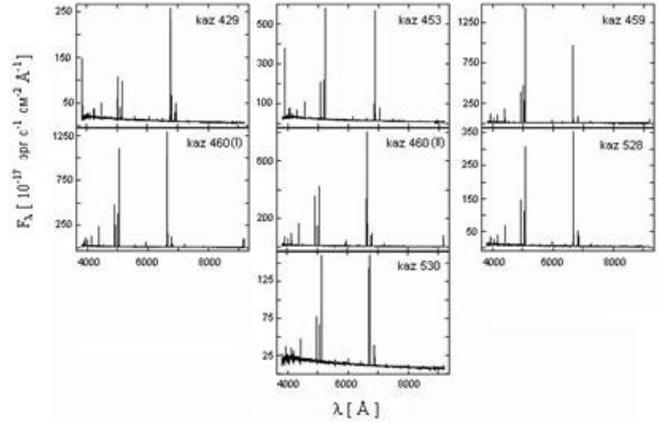

Fig. 2. Spectra of the HII regions in the Kazarian galaxies from SDSS DR5.

The program IRAF/SPLOT was used to analyze the spectra. The nonlinear wavelength scale of the extracted spectra was converted to a linear scale. The line intensities are fitted with a Gaussian profile, and the continuum is visually defined on both sides of each line. The SDSS spectra cover the range from λλ 3800 -9200 A, so the [OIII]λ3727 line does not lie within this range for galaxies with z < 0.025. In many of the SDSS spectra some of the peaks of the strongest emission lines, most often Hα and [OIIIλ]λ5007 are clipped. Thus, the theoretical values 2.85xI(Hβ) and 3x[OIII]λ4959, respectively, are taken for these lines.

The line intensities were corrected for reddening and for absorption in the hydrogen lines using the formula [17]

$$\frac{I(\lambda)}{I(H_\beta)} = \frac{F(\lambda)}{F(H_\beta)} \frac{EW_e(\lambda)+EW_a(\lambda)}{EW_e(\lambda)} \frac{EW_e(H_\beta)}{EW_e(H_\beta)+EW_a(H_\beta)} 10^{[C(H_\beta)f(\lambda)]}$$

where the I(λ) are the corrected and the F(λ) are the observed line intensities $EW_e(\lambda)$ and $EW_a(\lambda)$, respectively, are the equivalent widths of the observed emission line and of its component in absorption, and $f(\lambda)$ is the reddening function normalized to Hβ taken from Ref. 18. The theoretical ratios of the hydrogen lines I(λ)/I(Hβ) are taken from Ref. 19. Since strong lines may be clipped in some cases (In Table 2 the observed ratio F(Hα)/F(Hβ) is almost always less than 2.85.), $EW_a(\lambda)$ and C(Hβ) have been determined using only the ratios F(Hγ)/F(Hβ) and F(Hδ)/F(Hβ). Table 2 gives the values of $EW_e(\lambda)$, $EW_a(\lambda)$, C(Hβ), F(λ), and I(λ).

TABLE 2. Relative Intensities of Emission Lines in the HII Regions

|  | Kaz 429 | | Kaz 453 | | Kaz 459 | | Kaz 460(I) | | Kaz 460(II) | | Kaz528 | | Kaz 530 | |
|---|---|---|---|---|---|---|---|---|---|---|---|---|---|---|
|  | F($\lambda$) | I($\lambda$) | F($\lambda$) | I($\lambda$) | F($\lambda$) | I($\lambda$) | F($\lambda$) | I($\lambda$) | F($\lambda$) | I($\lambda$) | F($\lambda$) | I($\lambda$) | F($\lambda$) | I($\lambda$) |
| 3727 [O II] | 0.950 | 1.042 | 1.364 | 1.460 |  |  |  |  |  |  |  |  |  |  |
| 3868 [Ne III] | 0.066 | 0.071 | 0.249 | 0.264 | 0.270 | 0.283 | 0.143 | 0.143 | 0.058 | 0.058 | 0.201 | 0.212 | 0.284 | 0.301 |
| 4101 H$\delta$ | 0.200 | 0.260 | 0.208 | 0.260 | 0.239 | 0.259 | 0.247 | 0.247 | 0.248 | 0.248 | 0.219 | 0.263 | 0.217 | 0.243 |
| 4340 H$\gamma$ | 0.414 | 0.460 | 0.413 | 0.451 | 0.440 | 0.457 | 0.457 | 0.457 | 0.463 | 0.463 | 0.427 | 0.458 | 0.430 | 0.448 |
| 4363 [O III] | 0.008 | 0.008 | 0.032 | 0.032 | 0.038 | 0.039 | 0.011 | 0.011 | 0.009 | 0.009 | 0.018 | 0.018 | 0.021 | 0.022 |
| 4471 He I | 0.039 | 0.040 | 0.038 | 0.039 | 0.037 | 0.037 | 0.037 | 0.037 | 0.036 | 0.036 | 0.036 | 0.037 | 0.037 | 0.038 |
| 4686 He II |  |  | 0.014 | 0.014 | 0.007 | 0.007 | 0.006 | 0.006 |  |  |  |  |  |  |
| 4861 H$\beta$ | 1.000 | 1.000 | 1.000 | 1.000 | 1.000 | 1.000 | 1.000 | 1.000 | 1.000 | 1.000 | 1.000 | 1.000 | 1.000 | 1.000 |
| 4959 [O III] | 0.281 | 0.279 | 1.014 | 1.009 | 1.237 | 1.232 | 0.760 | 0.760 | 0.430 | 0.430 | 0.755 | 0.752 | 0.887 | 0.883 |
| 5007 [O III] | 0.884 | 0.875 | 2.961 | 2.940 | 3.714 | 3.690 | 2.271 | 2.271 | 1.266 | 1.266 | 2.254 | 2.240 | 2.640 | 2.620 |
| 5876 He I | 0.103 | 0.096 | 0.105 | 0.100 | 0.107 | 0.103 | 0.109 | 0.109 | 0.106 | 0.106 | 0.103 | 0.098 | 0.103 | 0.101 |
| 6312 [S III] |  |  | 0.011 | 0.010 | 0.013 | 0.012 | 0.010 | 0.010 | 0.009 | 0.009 | 0.007 | 0.007 | 0.007 | 0.007 |
| 6548 [N II] | 0.214 | 0.193 | 0.087 | 0.080 | 0.047 | 0.044 | 0.107 | 0.107 | 0.159 | 0.159 | 0.093 | 0.086 | 0.101 | 0.093 |
| 6563 H$\alpha$ | 2.765 | 2.850 | 2.872 | 2.850 | 2.749 | 2.850 | 2.668 | 2.850 | 2.586 | 2.850 | 2.605 | 2.850 | 2.739 | 2.850 |
| 6584 [N II] | 0.630 | 0.568 | 0.228 | 0.211 | 0.135 | 0.126 | 0.319 | 0.319 | 0.480 | 0.480 | 0.277 | 0.258 | 0.304 | 0.281 |
| 6678 He I | 0.031 | 0.027 | 0.031 | 0.028 | 0.031 | 0.028 | 0.031 | 0.031 | 0.028 | 0.028 | 0.033 | 0.031 |  |  |
| 6717 [S II] | 0.431 | 0.386 | 0.356 | 0.327 | 0.210 | 0.196 | 0.241 | 0.241 | 0.283 | 0.283 | 0.259 | 0.236 | 0.365 | 0.328 |
| 6731 [S II] | 0.300 | 0.268 | 0.251 | 0.231 | 0.152 | 0.142 | 0.177 | 0.177 | 0.206 | 0.206 | 0.179 | 0.163 | 0.247 | 0.222 |
| 7136 [Ar III] | 0.040 | 0.036 | 0.069 | 0.062 | 0.066 | 0.061 | 0.081 | 0.081 | 0.060 | 0.060 | 0.062 | 0.056 | 0.058 | 0.053 |
| 7320 [O II] | 0.013 | 0.011 | 0.027 | 0.024 | 0.023 | 0.021 | 0.017 | 0.017 | 0.012 | 0.012 | 0.021 | 0.019 | 0.021 | 0.019 |
| 7330 [O II] | 0.007 | 0.006 | 0.019 | 0.017 | 0.019 | 0.017 | 0.015 | 0.015 | 0.009 | 0.009 | 0.015 | 0.013 | 0.017 | 0.015 |
|  |  |  |  |  |  |  |  |  |  |  |  |  |  |  |
| C(H$\beta$) (dex) | 0.13 | | 0.10 | | 0.08 | | 0.00 | | 0.00 | | 0.09 | | 0.10 | |
| EW(H$\beta$) (Å) | 20.6 | | 28.2 | | 102.5 | | 108.1 | | 91.6 | | 49.5 | | 13.6 | |
| EW(abs) (Å) | 0.6 | | 0.8 | | 0.9 | | 0.0 | | 0.0 | | 1.3 | | 0.2 | |

## 3. Abundance of chemical elements

**3.1. Abundance of heavy elements.** In determining the abundance of ions, the differences in the electron temperatures in the zones where they are formed have been taken into account.

The electron temperature in the O$^{++}$ zones was calculated from an iterative procedure, using the following equation [20]:

$$t = \frac{1.432}{Log[(\lambda 4959 + \lambda 5007)/\lambda 4363] - LogC_T},$$

where

$$t = 10^{-4} T_e(OIII), \quad C_T = (8.44 - 1.09t + 0.5t^2 - 0.08t^3)\frac{1 + 0.0004x}{1 + 0.044x}, \quad x = 10^{-4} N_e t^{-0.5}$$

The electron density n_e([SII]) was determined using the ratio [SII]λ6717/[SII]λ6731. (The transition probabilities for these lines were taken from Ref. 21 and the collisional strengths, from Ref. 22.) The resulting values of n_e([SII]) were <150 cm$^{-3}$ in all cases. (See Table 3.)

TABLE 3. Elemental and Ion Abundances in the HII Regions

| Parameters | Kaz429 | Kaz429[1] | Kaz453 | Kaz459 | Kaz460(I) | Kaz460(II) | Kaz460(II)[1] | Kaz528 | Kaz 530 |
|---|---|---|---|---|---|---|---|---|---|
| Te ([O III]) (K) | 11020 | 9510 | 11841 | 11747 | 9190 | 10279 | 7830 | 10697 | 10848 |
| Te ([O II]) (K) | 10917 | 9617 | 11692 | 11607 | 9461 | 10147 | 9441 | 10589 | 10744 |
| Te ([S III]) (K) | 9758 | 9543 | 10904 | 10776 | 9306 | 8658 | 8666 | 9285 | 9508 |
| n_e ([S II]) (cm$^{-3}$) | 60 | 60 | 84 | 123 | 128 | 116 | 115 | 51 | 23 |
| O$^+$/H$^+$ ( x 10$^5$ ) | 2.46 | 5.31 | 4.06 | 3.91 | 11.12 | 4.69 | 7.40 | 5.53 | 5.38 |
| (O$^+$/H$^+$)$_{3727}$ ( x 10$^5$ ) | 2.91 | 5.00 | 3.13 | | | | | | |
| O$^{++}$/H$^+$ ( x 10$^5$ ) | 2.28 | 3.77 | 6.25 | 7.98 | 11.22 | 4.22 | 11.87 | 6.52 | 7.29 |
| O$^{+++}$/H$^+$ ( x 10$^7$ ) | | | 8.36 | 4.81 | 7.40 | | | | |
| O/H ( x 10$^5$ ) | 5.20 | 8.77 | 9.39 | 11.95 | 22.42 | 8.91 | 19.27 | 12.05 | 12.67 |
| 12+log(O/H) | 7.71 | 7.94 | 7.97 | 8.07 | 8.35 | 7.95 | 8.28 | 8.08 | 8.10 |
| N$^+$/H$^+$ ( x 10$^6$ ) | 8.71 | 12.14 | 2.82 | 1.68 | 7.12 | 8.84 | 10.70 | 4.25 | 4.45 |
| ICF | 1.93 | 1.69 | 2.56 | 3.04 | 2.39 | 1.90 | 2.62 | 2.19 | 2.37 |
| N/H ( x 10$^6$ ) | 16.86 | 20.57 | 7.22 | 5.10 | 17.09 | 16.80 | 28.15 | 9.35 | 10.58 |
| 12+log(N/H) | 7.23 | 7.31 | 6.85 | 6.70 | 7.24 | 7.22 | 7.45 | 6.97 | 7.02 |
| log(N/O) | -0.49 | -0.63 | -1.11 | -1.37 | -1.12 | -0.72 | -0.83 | -1.11 | -1.08 |
| Ne$^{++}$/H$^+$ ( x 10$^6$ ) | 5.30 | 9.65 | 15.15 | 16.70 | 22.59 | 5.69 | 19.77 | 17.77 | 23.90 |
| ICF | 1.22 | 1.26 | 1.17 | 1.14 | 1.71 | 1.23 | 1.16 | 1.20 | 1.18 |
| Ne/H ( x 10$^6$ ) | 6.50 | 12.14 | 17.74 | 19.07 | 38.80 | 7.03 | 23.06 | 21.35 | 28.30 |
| 12+log(Ne/H) | 6.81 | 7.08 | 7.25 | 7.28 | 7.58 | 6.84 | 7.36 | 7.33 | 7.45 |
| log(Ne/O) | -0.86 | -0.87 | -0.72 | -0.80 | -0.76 | -1.10 | -0.92 | -0.75 | -0.65 |
| S$^+$/H$^+$ ( x 10$^6$ ) | 1.19 | 1.63 | 0.88 | 0.54 | 1.09 | 1.07 | 1.28 | 0.78 | 1.04 |
| S$^{++}$/H$^+$ ( x 10$^6$ ) | | | 1.59 | 2.00 | 3.15 | 4.01 | 3.99 | 2.07 | 1.85 |
| ICF | | | 1.07 | 1.09 | 1.00 | 1.05 | 1.07 | 1.05 | 1.06 |
| S/H ( x 10$^6$ ) | | | 2.63 | 2.76 | 4.27 | 5.33 | 5.64 | 3.01 | 3.06 |
| 12+log(S/H) | | | 6.42 | 6.44 | 6.63 | 6.73 | 6.75 | 6.47 | 6.48 |
| log(S/O) | | | -1.60 | -1.64 | -1.72 | -1.22 | -1.53 | -1.60 | -1.61 |
| Ar$^{++}$/H$^+$ ( x 10$^7$ ) | 3.22 | 3.40 | 4.26 | 4.31 | 8.17 | 7.31 | 7.29 | 5.68 | 5.03 |
| ICF | 1.07 | 1.08 | 1.07 | 1.08 | 1.10 | 1.08 | 1.07 | 1.07 | 1.07 |
| Ar/H ( x 10$^7$ ) | 3.47 | 3.70 | 4.58 | 4.66 | 9.02 | 7.90 | 7.84 | 6.10 | 5.40 |
| 12+log(Ar/H) | 5.54 | 5.57 | 5.66 | 5.67 | 5.95 | 5.90 | 5.89 | 6.78 | 5.73 |
| log(Ar/O) | -2.13 | -2.39 | -2.36 | -2.41 | -2.39 | -2.05 | -2.39 | -2.29 | -2.37 |
| y$^+$ (average) | 0.0751 | 0.0730 | 0.0774 | 0.0767 | 0.0770 | 0.0747 | 0.0700 | 0.0731 | 0.0759 |
| y$^{++}$ (4686) | | | 0.0012 | 0.0006 | 0.0005 | | | | |
| η | | | 1.168 | 1.796 | 2.851 | 4.148 | 1.930 | 2.350 | 2.130 |
| ICF(He) | 1.000 | 1.000 | 1.007 | 1.012 | 1.022 | 1.038 | 1.013 | 1.017 | 1.015 |
| y = y$^+$ + y$^{++}$ | 0.075 | 0.073 | 0.078 | 0.078 | 0.079 | 0.078 | 0.071 | 0.074 | 0.077 |

[1]Values obtained by the Te(Ar/O) method

The electron temperatures for the ions OII, NeIII, SII, SIII, NII, ArIII, NeIII and HeII were determined using the relation between Te([OIII]) and the electron temperatures of these ions from Ref. 3. The electron temperatures for these ions in the HII regions are given in Table 3.

The following lines were used to determine the element abundances: [OII]λ3727 (or [OII]λλ7320,7331, if the [OII]λ3727 line was not observed (for galaxies with z < 0.025)) and [OIII]λλ4959,5007 for oxygen; [NII]λλ6548,6584 for nitrogen; [NeIII]λ3868 for neon; [SIII]λ6312 and [SII]λλ6717,6731 for sulfur; and, the line [ArIII]λ7135 for argon. The formulas from Ref. 3 were used to calculate the abundances of these ions.

The oxygen abundance was calculated according to $O/H=O^+/H^+ + O2^+/H^+$, except for three HII regions, in which the HeIIλ4686 line was observed. The fraction of $O3^+$ ions [3] was also taken into account to determine the total oxygen abundance in these three HII regions. (In all three cases, the fraction of $O3^+$ ions was <1% of the total oxygen content.) The total amount of the other elements was calculated according to $\frac{X}{H} = ICF(X)\frac{X^{+i}}{H^+}$. The ICFs for the heavy elements were determined using the formulas given in Ref. 3.

Table 3 lists the abundances of the heavy elements and their ions, the ICF, Te, and $n_e$ in the HII regions being studied.

Figure 3 shows plots of the ratios N/O, S/O, Ne/O, and Ar/O as functions of 12+log(O/H). The solid circles denote two low-excitation HII regions for which [OIII]λ4959/Hβ < 0.7 . Here it can be seen that these two regions differ from the other five (for which the ratios are indicated by open circles) in the values of all four ratios. This kind of difference has been observed previously [2], where it was suggested that the difference may be caused by an inaccurate determination of the intensity of the highly excited [OIII]λ4363 line in the low-excitation HII regions (where this line is faint). However, it was noted that in these low-excitation HII regions there may be an additional heating mechanism besides the radiation from young stars. In these two HII regions, I([OIII]λ4363) < 0.01×I(Hβ) (See Table 2) and the inaccuracy in determining the intensity of the [OIII]λ4363 line may be large. Note that these two objects lie relatively closer to the centers of their parent galaxies than to the others. (See Table 1.)

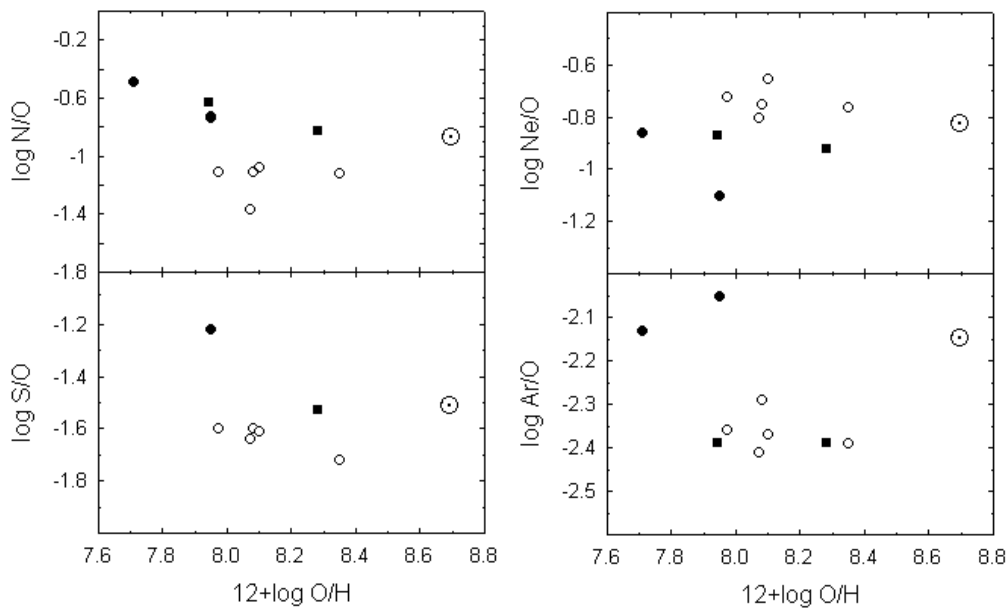

Fig. 3. log(N/O), log(Ne/O), log(S/O), and log(Ar/O) as functions of 12+log(O/H) for the HII regions. The values of these ratios for the sun [24] are indicated by the symbol ☉.

As an alternative, a method introduced in Ref. 2 was used to determine the electron temperature in these HII regions. Assuming that log(Ar/O) = –2.39 [3], the corresponding electron temperature was determined and it was used to determine the abundances of the other elements (the Te(Ar/O) method). The choice of the ratio log(Ar/O) for determining Te is more appropriate than log(S/O) or log(Ne/O), since the last two ratios depend more weakly on Te than does log(Ar/O) [2]. It can be seen from Table 3 and Fig. 3 that the ratios log(S/O) and log(Ne/O) (indicated by solid squares) obtained by the Te(Ar/O) method are close to the values for the other five regions, while the value of log(N/O) ended up smaller, although it was still larger than the value for the other five.

The averages of log(Ar/O), log(S/O), and log(Ne/O) (for all the HII regions) are -2.37, -1.62, and -0.78, respectively. These values are close to the averages of the corresponding rations obtained for high-excitation HII regions in Ref. 3, but differ from the averages of these ratios for HII regions obtained in Ref. 23. The differences in the values of log(Ar/O), log(S/O), and log(Ne/O) in this paper and in Ref. 23 may be caused by a difference in the method of determining these ratios. The [OIII]λ4363 line was observed in only 9% of the HII regions in the overall sample of Ref. 23, and the abundance of the chemical elements was determined empirically. These differences are greater for the ratio log(N/O) because it is more sensitive to Te than the others (log(Ar/O), log(S/O), and log(Ne/O)) [2]. The above mentioned ratios for the sun are -2.14, -1.5, and -0.82, respectively [24]. The values of log(Ar/O) and log(S/O) obtained here are different from the corresponding values for the sun. These differences may be caused by the inaccuracies in the determinations of these ratios both for the sun [3] and in our work.

The log (N/O) abundance ratio of investigated HII regions is in the interval -0.63 ÷ -1.37. For a given metallicity, the value of log(N/O) for these HII regions is greater than for the HII regions studied in Ref. 23. In a plot of log(N/O) vs 12+log(O/H), the HII regions studied here lie in the same area as the high-excitation HII regions in Refs. 2 and 3. This suggests similar star formation histories and evolutionary status for theme. Most likely, the ages of investigated HII regions are larger than 100-300 Myr, required for the enrichment in nitrogen by intermediate-mass stars [2,3,12].

**3.2. The helium abundance**. The abundance of helium was determined from the HeIλ4471, HeIλ5876, and HeIλ6678 lines (for He$^+$) and HeIIλ4686 (from He$_{++}$) using the formulas given in Refs. 17 and 25.

The total amount of helium was calculated according to He/H=ICF(He)y, where y = y$^+$ + y$^{2+}$ ≡ He$^+$/H + He$^{++}$/H and ICF(He) = η (0.005 + 0.001η) [26], with $\eta = \frac{O^+}{S^+} \frac{S^{2+}}{O^{2+}}$. The correction factor ICF(He) takes the neutral helium content into account (although the fraction of neutral helium is larger, (≥ 5%), when η ≥ 10 [26], while η < 5 in the HII regions being studied; see Table 3). The [SIII]λ6312 emission line was not observed in Kaz 429, so that ICF(He) = 1 for it.

The values of y$^+$ (as the average of y$^+$(4471), y$^+$(5876) and y$^+$(6678)), y$^{2+}$, η, ICF(He), and y are listed in Table 3.

## 4. Stellar population

The hydrogen lines effectively re-emit the integrated stellar luminosity of galaxies shortward of the Lyman limit, so they provide a direct, sensitive probe of the young massive stellar population. These lines are used for the determination the quantity of ionizing stars and star formation rate in investigated HII regions.

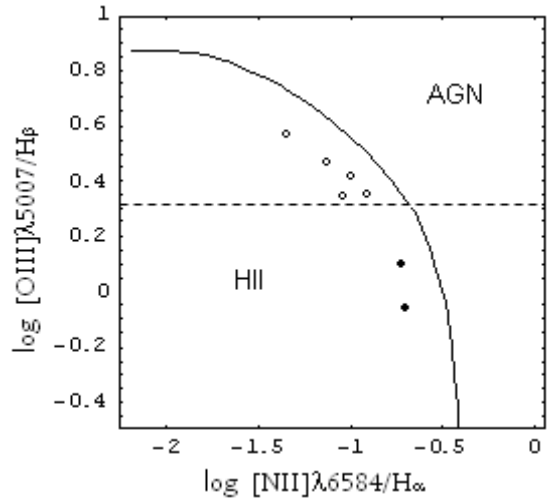

Fig. 4. Diagnostic diagram [OIII]$\lambda$5007/H$\beta$ vs [NII]$\lambda$6583/H$\alpha$. Filled circles are the HII regions with [OIII] $\lambda$4959/H$\beta$ < 0.7 and open circles are the HII regions with [OIII] $\lambda$4959/H$\beta$ $\geq$ 0.7. The filled line separates the HII regions from galaxies ionized by a non-thermal radiation (labeled "AGN") and the dashed line corresponds to [OIII]$\lambda$4959/H$\beta$=0.7.

Figure 4 shows the classical diagnostic diagram [OIII]$\lambda$5007/H$\beta$ vs [NII]$\lambda$6583/H$\alpha$ used to distinguish objects ionized by massive main sequence stars from objects ionized by non-thermal radiation. The number of ionizing stars in an HII region can be estimated from the luminosity in the H$_\beta$ emission line The luminosity L(H$\beta$) can be determined from F(H$\beta$) — L(H$\beta$)=1.2 $10^{50}$D$^2$F(H$\beta$), where F(H$\beta$) is the H$\beta$ line flux (corrected for absorption) and D is the distance to the object in Mpc. If the optical thickness of the region beyond the Lyman series limit is considerably greater than unity ($\tau_c \gg 1$), then for Te = $10^4$K the number of photons in the Lyman continuum is NLyC=2.06 $10^{12}$L(H$\beta$) [28]. Assuming that the LyC photons are emitted by stars of type O7 V (in Ref. 29, the ionizing stars in Kaz 460(I) and Kaz 460(II) were found to be of types O5 and O8, respectively), we obtain the number of these stars in the HII regions under study. (The amount of LyC emitted by O7 V stars was taken from Ref. 30.) Note that the number of stars was computed for a central part of the HII regions with an angular radius of 1".5 (the observed SDSS spectra correspond to an aperture of this size).

Table 4 gives the values of F(H$\beta$), L(H$\beta$), D, NLyC, R – the linear radius corresponding to 1".5, r – the linear radius of the entire HII region, and N(O7 V) the number of O7 V stars in the HII region within an angular radius of 1".5. Here the differences between the sizes of the entire HII region (r) and of the region corresponding to 1".5 (R) are comparatively greater for Kaz 459, Kaz 460(I), and Kaz 460(II) (their redshifts are smaller than those of the others). That is, we have obtained a lower limit for the number of O7 V stars in them. The number of O7 V stars in the HII regions lies between $0.45 \times 10^3 \div 7.5 \times 10^3$. The largest number of these stars was found in the giant HII region in the galaxy Kaz 453.

TABLE 4. The Number of Ionizing Stars in the HII Regions

| Parameters | Kaz 429 | Kaz 453 | Kaz 459 | Kaz 460(I) | Kaz 460(II) | Kaz528 | Kaz 530 |
|---|---|---|---|---|---|---|---|
| D (Mpc) | 121.6 | 188.4 | 51.2 | 42 | 42 | 71.2 | 88.4 |
| R (pc) | 885 | 1370 | 370 | 305 | 305 | 517.5 | 640 |
| r (pc) | 885 | 1351 | 459 | 437 | 332 | 507 | 595 |
| F(Hβ) ( x $10^{-17}$) (erg $s^{-1}$ $cm^{-2}$) | 398 | 861 | 1498 | 1575 | 1031 | 527 | 238 |
| L (Hβ) ( x $10^{39}$) (erg $s^{-1}$) | 7.06 | 36.68 | 4.71 | 3.33 | 2.18 | 3.20 | 2.23 |
| NLyC ( x $10^{51}$) | 14.55 | 75.56 | 9.71 | 6.87 | 4.50 | 6.60 | 4.60 |
| N (O7 V) | 1455 | 7556 | 971 | 687 | 450 | 660 | 460 |
| SFR ( x $10^2$) ($M_\odot$ $year^{-1}$) | 15.71 | 81.60 | 10.48 | 7.42 | 4.86 | 7.13 | 4.97 |

The values of L(Hβ) were used to determine the star formation rate (SFR) [31]. SFR for these HII objects lies within the range $0.05 \div 0.81$ $M_\odot$ $year^{-1}$. Table 4 shows that the star formation rate, like the number of O7 V stars, is greatest in the giant HII region of the irregular galaxy Kaz 453. The luminosity L(Hβ) and, therefore, the star formation rate, are one order as in the HII regions of spiral and irregular galaxies [32].

## 5. Conclusion

The results of this study of seven HII regions in Kazarian galaxies can be summarized as follows:

1. Among investigated HII regions, objects with extremely low metallicity are not found (12 + log(O/H) < 7.6, i.e. $Z < Z_\odot/12$). The oxygen abundance 12+log(O/H) lies in the range from ~7.94 to ~8.35. The mean log(S/O), log(Ar/O) and log(Ne/O) abundance ratios are equal to: -1.63, -2.37 and -0.78, respectively. These mean values are close to them derived for high-excitation HII regions [2,3].

2. The log (N/O) abundance ratio of investigated HII regions is in the interval $-0.63 \div -1.37$. They occupy the same area in the diagram N/O - O/H as the high-excitation HII regions [2,3]. This suggests similar star formation histories and evolutionary status for theme. Most likely, the ages of investigated HII regions are larger than 100-300 Myr, required for the enrichment in nitrogen by intermediate-mass stars

3. The quantity of ionizing stars is in interval $0.45 \cdot 10^3 \div 7.5 \cdot 10^3$. The star formation rate is one order as in HII regions in spiral and irregular galaxies, and is in the interval $0.05 \div 0.81$ $M_\odot year^{-1}$. The SFR and quantity of O7 V stars are largest in giant HII region of irregular galaxy Kaz 453.

The author thanks *A.R. Petrosian* and *M.Z. Kazarian* for helpful discussions and valuable comments.

Data from the Lyon-Medon database of extragalactic objects (HyperLeda), supported by the LEDA group and the CRAL observatory in Lyon, and from the fifth database of the Sloan Digital Sky Survey (SDSS DR5), which is available for open access at http://www.sdss.org/dr5, have been used in this paper. Funding for the SDSS and SDSS-II was provided by the Alfred P. Sloan Foundation, the Participating Institutions, the National Science Foundation, the U.S. Department of Energy, the National Aeronautics and Space Administration, the Japanese Monbukagakusho, the Max Planck Society, and the Higher Education Funding Council for England. The SDSS was managed by the Astrophysical Research Consortium for the Participating Institutions. For image processing the ADHOC software (www.astrsp-mrs.fr/index_lam.html) developed by Dr. Jacques Boulesteix (boulesteix@observatoire.cnrs-mrs.fr; Marseille Observatory, France) was in intensive use.